%% file: main.tex
\documentclass[10pt, conference]{IEEEtran}

\usepackage{url}
\usepackage[sort]{cite}
\usepackage[usenames, dvipsnames]{color}
\usepackage[nolist,nohyperlinks]{acronym}
\usepackage[table,xcdraw]{xcolor}
\usepackage{tabularx}
\usepackage[pdftex]{graphicx}
\usepackage{multirow}
\graphicspath{{images/}}
\DeclareGraphicsExtensions{.pdf,.jpeg,.png}
\usepackage{array}
\usepackage{caption}
\usepackage{cleveref}
\usepackage{subcaption}
\crefname{section}{§}{§§}
\crefformat{section}{\S#2#1#3}

\makeatletter
\def\ps@IEEEtitlepagestyle{%
  \def\@oddfoot{\mycopyrightnotice}%
  \def\@evenfoot{}%
}
\def\mycopyrightnotice{%
    \begin{minipage}{\textwidth}
        \centering\tiny{DOI 10.1109/NETSOFT.2018.8460125 \copyright 2018 IEEE.  Personal use of this material is permitted.  Permission from IEEE must be obtained for all other uses, in any current or future media, including reprinting/republishing this material for advertising or promotional purposes, creating new collective works, for resale or redistribution to servers or lists, or reuse of any copyrighted component of this work in other works.}
    \end{minipage}
  \gdef\mycopyrightnotice{}
}

\begin{document}

\title{Evolving SDN for Low-Power IoT Networks}

\author{
\IEEEauthorblockN{
Michael Baddeley\IEEEauthorrefmark{1}, 
Reza Nejabati\IEEEauthorrefmark{1}, 
George Oikonomou\IEEEauthorrefmark{1},
Mahesh Sooriyabandara\IEEEauthorrefmark{2},
Dimitra Simeonidou\IEEEauthorrefmark{1}}

\IEEEauthorblockA{\IEEEauthorrefmark{1}High Performance Networks Group, University of Bristol, Bristol, United Kingdom, \\ Email: \{m.baddeley, reza.nejabati, g.oikonomou, dimitra.simeonidou\}@bristol.ac.uk}
\IEEEauthorblockA{\IEEEauthorrefmark{2}Toshiba Research Europe Ltd., Bristol, United Kingdom, Email: mahesh@toshiba-trel.com }
}

\maketitle

\begin{abstract}
Software Defined Networking (SDN) offers a flexible and scalable architecture that abstracts decision making away from individual devices and provides a programmable network platform. Low-power wireless Internet of Things (IoT) networks, where multi-tenant and multi-application architectures require scalable and configurable solutions, are ideally placed to capitalize on this research. However, implementing a centralized SDN architecture within the constraints of a low-power wireless network faces considerable challenges. Not only is controller traffic subject to jitter due to unreliable links and network contention, but the overhead generated by SDN can severely affect the performance of other traffic. This paper addresses the challenge of bringing high-overhead SDN architecture to IEEE 802.15.4 networks. We explore how the traditional view of SDN needs to evolve in order to overcome the constraints of low-power wireless networks, and discuss protocol and architectural optimizations necessary to reduce SDN control overhead - the main barrier to successful implementation. Additionally, we argue that interoperability with the existing protocol stack is necessary to provide a platform for controller discovery, and coexistence with legacy networks. We consequently introduce $\mu$SDN, a lightweight SDN framework for Contiki OS with both IPv6 and underlying routing protocol interoperability, as well as optimizing a number of elements within the SDN architecture to reduce control overhead to practical levels. We evaluate $\mu$SDN in terms of latency, energy, and packet delivery. Through this evaluation we show how the cost of SDN control overhead (both bootstrapping and management) can be reduced to a point where comparable performance and scalability is achieved against an IEEE 802.15.4-2012 RPL-based network. Additionally, we demonstrate $\mu$SDN \footnote{Available: https://github.com/mbaddeley/usdn} through simulation: providing a use-case where the SDN configurability can be used to provide Quality of Service (QoS) for critical network flows experiencing interference, and we achieve considerable reductions in delay and jitter in comparison to a scenario without SDN. 

\end{abstract}


\input{./sec_01_intro}
\input{./sec_02_background}

\input{./sec_03_approach}

\input{./sec_04_results}

\input{./sec_05_conclusion}

\section*{Acknowledgements}

The authors wish to acknowledge the financial support of the Engineering and Physical Sciences Research Council (EPSRC) Centre for Doctoral Training (CDT) in Communications (EP/I028153/1), as well as Toshiba Research Europe Ltd. 

\bibliographystyle{IEEEtran}
\bibliography{main}

\end{document}

%% file: sec_01_intro.tex
\section{Introduction} \label{sec_intro}

\textbf{Context:} 
In recent years Software Defined Networking (SDN) has gained traction as a means of bringing scalability and programmability to network architecture. Particularly in Data Center and Optical Networks, SDN has been shown to offer a high degree of network configurability, reduction in capital expenditure, and a platform for virtualizing network functions \cite{sdn_comprehensive_survey}.

\textbf{Motivation:} 
The advantages of SDN have led to a number of research efforts to apply the concept within the IEEE 802.15.4 low-power wireless standard, which underpins many Internet of Things (IoT) and sensor networks. In particular, the reconfigurability conferred through SDN would allow low-power wireless networks to treat sensor and control traffic on a per-flow basis, providing guarantees to critical data whilst optimizing the network for low-energy communication. Elements of this idea can be seen in the approach to centralized scheduling defined within the IETF 6TiSCH architecture \cite{6tisch_ietf_architecture}, which uses SDN concepts to provide spatial and frequency diversity within IEEE 802.15.4-2015 industrial IoT networks.

\textbf{Challenge:}  
IoT sensor networks typically consist of constrained devices in a Low-Power and Lossy Network (LLN), and are limited in terms of reliability, throughput, and energy. Implementing a centralized SDN architecture in this environment therefore faces considerable challenges: not only is controller traffic subject to jitter due to unreliable links and network contention, but the overhead generated by SDN can severely affect the performance of other traffic. These limitations force us to revisit a number of traditionally held assumptions about how SDN operates, and how it can be applied within a constrained environment.

\textbf{Approach:} 
In this paper, we tackle the challenge of adapting high-overhead SDN architecture for constrained, low-power wireless networks. We introduce $\mu$SDN, a low-overhead SDN architecture which builds on recent trends towards centralization in protocols for IEEE 802.15.4 networks, and extends concepts introduced in recent works: where efforts have mainly considered non-IPv6 networks. $\mu$SDN implements additional optimization techniques, compatibility with IPv6 networks, and interoperability with existing distributed routing protocols such as RPL\cite{rpl_rfc} (Routing Protocol for Low-Power and Lossy Networks). We then use $\mu$SDN to evaluate the effect of SDN traffic network performance, and consider a scenario where SDN can improve Quality of Service (QoS) for high-priority flows over traditional approaches.

\textbf{Contribution:} 
\begin{itemize}
\item We introduce a number of optimization techniques to reduce control overhead and manage the challenges faced in applying SDN within low-power, multi-hop wireless networks.
\item We incorporate these techniques within $\mu$SDN, a lightweight SDN architecture for low-power wireless networks, and implement $\mu$SDN on top of the Contiki Operating System (OS) \cite{contiki} for constrained IoT networks. 
\item We evaluate the performance of $\mu$SDN, and show that $\mu$SDN maintains scalability when compared against a conventional IEEE 802.15.4-2012 network, whilst allowing the network to benefit from SDN architecture.
\item We present results showing how $\mu$SDN can maintain application and control paths on a per-flow basis, dramatically reducing delay and jitter in an interference scenario where conventional approaches struggle.
\end{itemize}

\textbf{Outline:} 
In Section \ref{sec_sdn} we examine, in general terms, the advantages inherent within SDN architecture, before discussing the motivation for applying SDN within low-power IoT scenarios in \ref{sec_motivation}. We then cover the challenges that must be overcome in order to apply SDN within a low-power, multi-hop environment in \ref{sec_challenges}. Additionally, in \ref{sec_rpl}, we provide necessary background information on RPL \cite{rpl_rfc}, the routing and topology control protocol typically employed within the IEEE 802.15.4 stack. An overview of key related works examining SDN in low-power wireless networks is given in \ref{sec_related_work}. In Section \ref{sec_approach} we frame the problem within the context of the challenges set out in \ref{sec_challenges}, covering various approaches necessary to optimize SDN architecture for constrained networks. We introduce $\mu$SDN, our lightweight SDN framework, in Section \ref{sec_design}, where we discuss its architectural design and implementation. Finally, we evaluate SDN overhead in IEEE 802.15.4 networks in Section \ref{sec_results}, as well as providing a use-case in which SDN reduces both delay and jitter for selected flows in an interference scenario. We then conclude in Section \ref{sec_conclusion}.

%% file: sec_02_background.tex
\section{Background and Related Work}
\label{sec_background}
\subsection{The Advantages of SDN}
\label{sec_sdn}
Though originally conceived for campus networks, SDN has been proposed as a solution to some of the problems inherent within traditional network architecture \cite{sdn_comprehensive_survey}. SDN separates the data and control planes by abstracting distributed state, network specification, and device forwarding \cite{shenker_future_past}. Through a network state model exposed by the Network Operating System (NOS), applications are then able to provide network services without knowledge of the underlying hardware or topology. The NOS utilizes data forwarding protocols, such as OpenFlow \cite{openflow}, to configure the network state based on compiled network behavior defined by an application layer. Unavailable with current protocols, SDN can provide a platform for virtualization of network functions and dynamic reconfiguration of services.

\textbf{Network (Re)Configurability:} In wired and optical networks, the programmability provided by SDN allows configuration of forwarding paths in the network, protocol independence, and customized processing of individual flows. In sensor networks in particular, this would allow freedom from protocol constraints. It would enable, for example, more stable networks to be configured to deliver greater performance, whereas more dynamic and mobile networks would be able to divert resources to critical and high-priority flows.

\textbf{Global Knowledge:} Constrained wireless networks typically employ distributed protocols in order to reduce the overall load on the network and minimize the inherent uncertainty. Whilst this approach has been widely successful, there is a growing acknowledgement that, through global knowledge, there are a number of areas in which centralized architectures could provide benefits to low-power wireless networks, particularly in managing interference and heterogeneity in dense and non-isolated networks.

\textbf{Virtualization:} The abstraction of protocol logic away from individual nodes lends itself well to the introduction of virtualization and network slicing. These are seen as key enablers in the provision of multi-tenant IoT networks, where multiple operators can utilize network infrastructure from a single vendor.

\subsection{Motivation for SDN in Low-Power IoT}
\label{sec_motivation}
SDN is now seen as a key enabler for next-generation wireless networks, particularly in low-power IoT sensor networks, which typically operate within an extremely constrained environment. Specifically, within IEEE 802.15.4-2012, power limitations force the use of multi-hop mesh networking in order to allow the network to reach beyond the radio transmission range. Typical networks may include dozens to hundreds of devices (within a single mesh) with multiple sensors per device. However, multiple networks may be connected across a backbone network, and protocols such as 6LoWPAN \cite{6lowpan_rfc} allow devices to be interoperable with IPv6. The flexibility and scalability provided by SDN presents further opportunity to move beyond the traditional notions of low-power IoT as small, horizontal islands serving a single application, typically categorized as one of three areas: \textit{Data Collection} (many-to-one), \textit{Alerts and Actuation} (point-to-point), and \textit{Data Dissemination} (one-to-many). The opportunity for SDN in this context can be summed up through examining the advantages in the previous section.

Using \textit{Global Knowledge}, SDN can help distribute flows and allocate network resources according to QoS requirements. This concept is already touched upon in the work on centralized scheduling for IEEE 802.15.4-2015 (as discussed in Section \ref{sec_related_work}).

\textit{Network (Re)configuration} can allow low-power wireless networks can be re-purposed on an ad-hoc basis, based on changing application and business needs. This would free low-power networks from serving a single application over their lifetime, and allow operators to add new sensors, actuators, and network capabilities with relative ease - without updating firmware.

By employing the SDN architecture to \textit{Virtualize} network functions such as routing, security, and data aggregation, IoT sensor networks can take advantage of greater computing resources at the controller. As well as allowing functions to be initialized or torn down depending on application needs, this process additionally permits the association of flows with individual functions. Moving from a horizontal island, SDN can allow the network to dynamically serve multiple applications, such as both \textit{Data Collection} and \textit{Actuation}, with varying QoS requirements.

\subsection{The Challenge of Low-Power Wireless Mesh}
\label{sec_challenges}
SDN is, by nature, an architecture with a high associated overhead: both in terms of the centralized control traffic, and also from flowtable lookups. Low-power wireless networks, on the other hand, are the antithesis of this. We provide an overview of the main constraints faced, and how this affects the task of trying to implement SDN within the network.

\textbf{Device Hardware Restrictions:} Low-power wireless networks consist of constrained devices with limited energy, memory, and processing capabilities. These restrictions allow devices to operate for months, or even years, on little power. The consequence of this is that concessions need to be made at all layers of the network stack. This is particularly limiting for SDN, which traditionally employs devices capable of processing thousands of flows per second, and sorting through tables that can sometimes hold hundreds of thousands of entries. Yet IEEE 802.15.4 devices often have only a few KB of memory, and excessive radio activity will quickly deplete a node's energy supply.

\textbf{Unreliable Links:} The lossy medium present in low-power wireless networks means they can be prone to unreliability. This is a direct consequence of the low-power hardware requirements, which forces concessions at the physical and MAC layers, but in order for SDN applications to provide effective decisions, there needs to be an up-to-date model of the network. The compounded problems of lossy links and a multi-hop mesh network means that addressing this can be problematic. Packets updating the controller of the network state can be dropped or subject to severe delays.

\textbf{Fragmentation:} IEEE 802.15.4 has a Maximum Transmission Unit (MTU) of 127B. After the link-layer header, the 6LoWPAN standard \cite{6lowpan_rfc} introduces IP capabilities but further reduces remaining space in a single, unfragmented packet. A full IPv6 6LoWPAN implementation with 64-bit addressing allows for a mere 53B of application data. In order to prevent packet fragmentation, and hence multiple transmissions per packet, SDN control messages need to fit within this allotted length.

\textbf{Interference:}
The low-power nature of transmissions means IEEE 802.15.4 networks can be sensitive to interference from nearby higher-power communications operating at the same frequency, such as IEEE 802.11 networks transmitting on the same channel at 2.4 GHz. This can potentially affect entire branches of the network, and hamper the delivery of messages from/to sensors and actuators. In an SDN architecture with centralized control, this prevents nodes from querying or receiving instructions from the controller.

\textbf{Multi-Hop Mesh Topology:} Distributed routing protocols, such as RPL, are commonly used to locally maintain topology whilst reducing control overhead in the overall network. As low-power devices reduce radio range, a multi-hop mesh allows networks to be extended over a greater area than if all nodes communicated with a single base station. Unfortunately, by introducing multiple hops, link uncertainty is compounded across the hop distance and can increase the chance of packets being dropped along the way.

\subsection{A Brief Overview of RPL} 
\label{sec_rpl}
RPL (Routing Protocol for Low-Power and Lossy Networks) \cite{rpl_rfc} forms an integral part of many low-power wireless networks. It allows the formation of tree-like ad-hoc network topologies, where each node keeps an immediate one-hop parent based on a configurable Objective Function (OF) that determines which parent to select (often the node's rank within the graph, or link metrics). As nodes form part of the topology based solely on information provided by their immediate neighbors, RPL is effective in allowing nodes to quickly send information up the tree. However, the RPL graph forces nodes nearer the root to serve messages from nodes further down the tree, and exacerbates energy loss in nodes nearer the root. Some RPL terminology is used in later sections, and a brief description of these terms is given below:

\begin{itemize}
\item \textit{Direction-Orientated Directed Acyclic Graph (DODAG):} A tree-like graph with no cycles, and single root node with no outgoing edge (although this often acts as a border router).
\item \textit{DODAG Information Solicitation (DIS):} ICMPv6 message used by nodes to request RPL DAG information from one-hop neighbors.
\item \textit{DODAG Information Object (DIO):} ICMPv6 message sent as a response to DIS messages.
\item \textit{Destination Advertisement Object (DAO):} Sent from child nodes to the parent or root (depending on the RPL mode) in order to advertise themselves as a destination within the DAG.
\item \textit{RPL Storing Mode:} Nodes maintain a routing table for their Sub-DODAG.
\item \textit{RPL Non-Storing (NS) Mode:} Nodes only know their parent, and the root keeps a routing table for the whole DODAG.
\end{itemize}

\subsection{Related Work}
\label{sec_related_work}
SDN is an increasingly well-defined concept which has been successfully applied to other networking areas. However, there is still considerable debate of what SDN means when it comes to low-power wireless networks. The IETF 6TiSCH Working Group (WG) \cite{6tisch_ietf_architecture} is engaged in efforts to develop scheduling mechanisms for IEEE 802.15.4-2015 TSCH, which allowed the creation of channel hopping schedules but did not define how these schedules should be configured or maintained. 6TiSCH aims to incorporate SDN concepts within the standard, but foregoes traditional SDN elements such as flowtables and focuses more on the centralized allocation of resources (the TSCH {channel/time} slots) within the network.

A number of more traditional SDN architectures for Low-Power Wireless Networks have been proposed and we briefly summarize these. In particular, we present the prevailing ideas from key contributions in this area; however, this is not an extensive list of works, which are covered in detail in recent surveys \cite{wsdn_survey_taxonomy, sdwn_opportunities_challenges,sdn_for_iot_survey}. 

Sensor OpenFlow \cite{sensor_openflow} were early advocates for using SDN in sensor networks. Their proposal highlights the difficulties of implementing Out-Of-Band (OOB) control plane communication within a sensor network, and explicitly argues for a \textit{custom low-power protocol}, rather than utilizing OpenFlow directly. They also propose \textit{energy saving through data aggregation}, and attempt to mitigate SDN overhead through the introduction of Control Message Quenching (CMQ) \cite{cmq}. This technique suppresses additional queries upon flowtable misses, purportedly allowing the network sufficient time to respond to the initial request.

Constanzo et al. \cite{sdwn} propose SDWN, an architectural framework which highlights novel uses for SDN in Low-Power Wireless Sensor Networks. The authors introduce the idea of using SDN flowtables to facilitate \textit{data aggregation} and \textit{Radio Duty-Cycling (RDC)} techniques, theoretically allowing SDN to programmatically configure the energy consumption of the node. In addition, to further reduce energy consumption, SDWN  suggests a form of \textit{Protocol Oblivious Forwarding (POF)} \cite{pof} to reduce memory footprint, allowing flowtables to match on byte arrays within the packet.

Following from the proposals in SDWN, the authors of SDN-WISE \cite{sdn_wise} provide a public implementation of the architecture using Contiki \cite{contiki}. SDN-WISE introduces \textit{stateful flowtables}, essentially turning the flowtables into a Finite State Machine (FSM). This allows simple controller logic to be `programmed' into the nodes, where they can perform certain actions under one state, whilst performing a different set of actions when in another. For example, this could be used to allow nodes to run their SDN flowtable actions in a low-energy mode.

More recent works in the field directly consider the overhead incurred by SDN in Low-Power Wireless Networks, and try to reduce the overhead of other protocols within the stack. CORAL \cite{coral-demo} takes a similar approach to this paper, in adapting SDN for IPv6 based IEEE 802.15.4 RPL networks, and tries to deal with network overhead by using SDN to fine-tune timer settings in the RPL in order to reduce the number of RPL transmissions after initialization and provide more resources for the SDN protocol.

%% file: sec_03_approach.tex
\section{Optimizing SDN for Low-Power Wireless}
\label{sec_approach}

Section \ref{sec_challenges} categorizes the challenges faced by SDN in low-power wireless networks. We address these challenges by looking at four core areas and how they might be optimized: the SDN \textit{protocol}, the SDN \textit{architecture}, the SDN \textit{flowtable and buffers}, and the SDN \textit{controller}. 

\textbf{Protocol Optimization:}
\begin{itemize}
\item \textit{Eliminate Fragmentation} through tailoring the SDN control protocol so that it doesn't exceed the allocated packet size after the link layer and 6LoWPAN headers.
\item \textit{Reduce Packet Frequency} to minimize potential for congestion, as well as reduce opportunities for retransmissions at the link layer
\item \textit{Match on Byte Arrays/Index} rather than specific header fields, allowing greater reconfigurability and programmability in the mesh.
\end{itemize}

\break
\textbf{Architectural Optimization:}
\begin{itemize}
\item \textit{Use Source Routing} to prevent intermediate nodes from generating new control requests as the packet is transported from source to destination (assuming that the intermediate nodes have no rules for that flow).
\item \textit{Throttle Control Messages} ensuring that repeated control requests, from a node to the controller, are not sent in quick succession. Additionally, this has security implications in that it is a possible defense against a denial-of-service style attack.
\item \textit{Refreshing Flowtable Timers} reduces reliance on instructions from the controller as repeated successful matches will not expire. This is, however, a trade-off between configurability and performance.
\end{itemize}

\textbf{Memory Optimization:}
\begin{itemize}
\item \textit{Re-Use Flowtable Matches/Actions} by eliminating repeated entries. For example, if there are entries for two flows which are then forwarded to the same destination, that forwarding action should be stored as a single item, rather than being included in both entries.
\item \textit{Reduce Buffer Sizes} by setting specific fields to be buffered at the initial controller configuration, rather than buffering the whole packet.
\end{itemize}

\textbf{Controller Optimization:}
\begin{itemize}
\item \textit{An Embedded Controller} would allow simple requests to be responded to more quickly, rather than sending them on to the external IPv6 backbone network.
\end{itemize}

\section{$\mu$SDN Design and Implementation}
\label{sec_design}

\subsection{Overview}
\begin{figure}[ht]
  \centering
  \includegraphics[width=0.7\columnwidth]{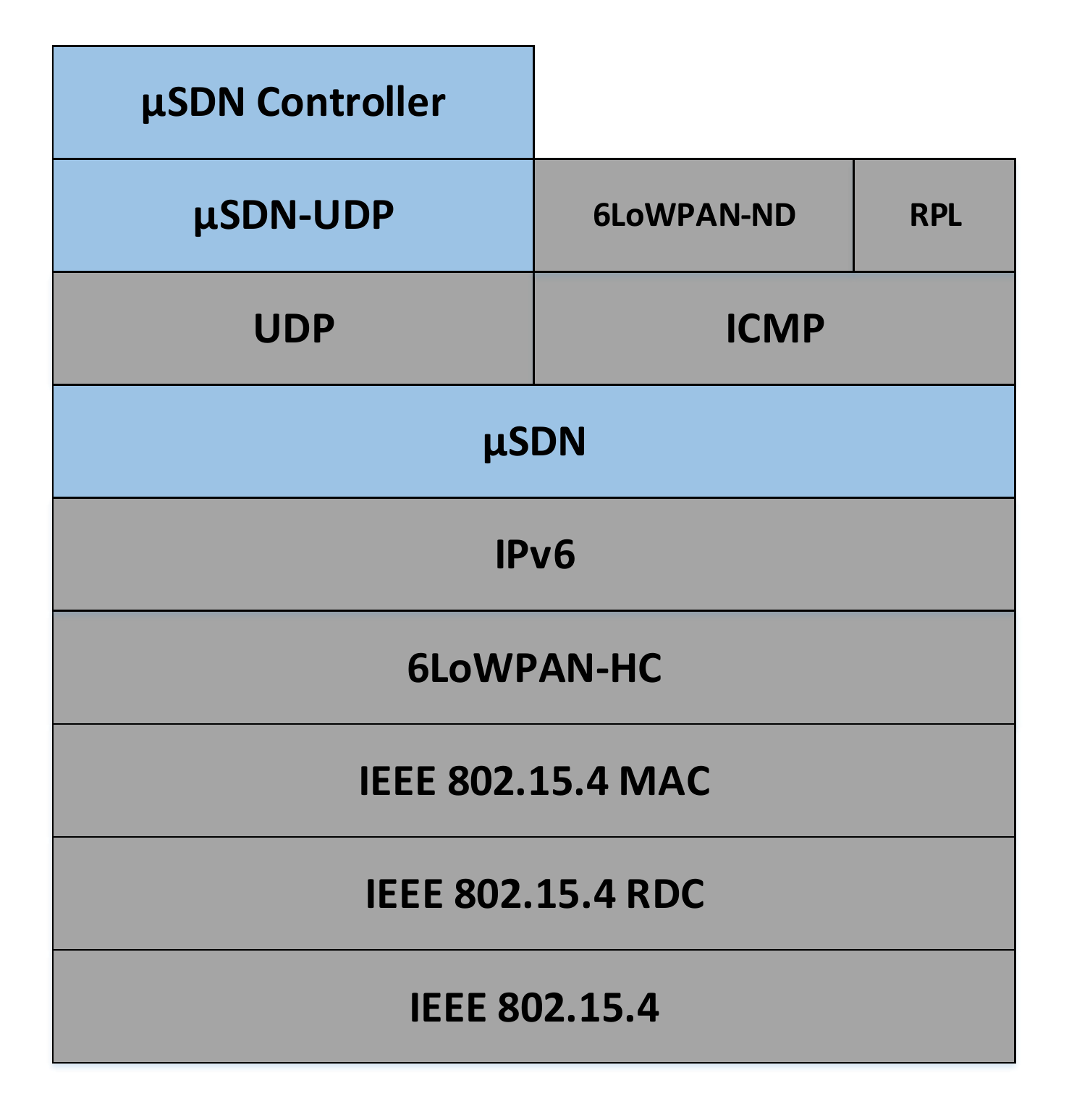}
  \caption{$\mu$SDN in IEEE 802.15.4-2012 stack with CSMA MAC Layer.}
\label{fig:usdn_stack}
\end{figure}

In order to provide a platform for SDN experimentation in wireless sensor networks we have implemented $\mu$SDN, a lightweight SDN framework for Contiki OS. $\mu$SDN builds on some of the architectural concepts proposed in the recent works highlighted in Section \ref{sec_related_work}, whilst incorporating the optimization techniques from Section \ref{sec_approach} in order to mitigate control overhead and enhance scalability. $\mu$SDN sits above the IP layer within the IEEE 802.15.4-12 stack (as shown in Figure \ref{fig:usdn_stack}), and $\mu$SDN nodes are, in theory, inter-operable with legacy nodes in a IEEE 802.15.4 network. Although left unexplored in this paper, the topic of incorporating SDN nodes within a wider low-power wireless network is a potential area for future work, as it could potentially provide an opportunity to use SDN nodes to locally control branches in a hierarchical manner, thus potentially reducing the overhead cost of SDN even further, and allowing local controllers to make decisions without navigating a large number of hops. Finally, $\mu$SDN utilizes the RPL to provide a fallback communication path to the controller, though this could be replaced with any distributed routing protocol. 

\begin{figure}[b]
  \centering
  \includegraphics[width=0.6\columnwidth]{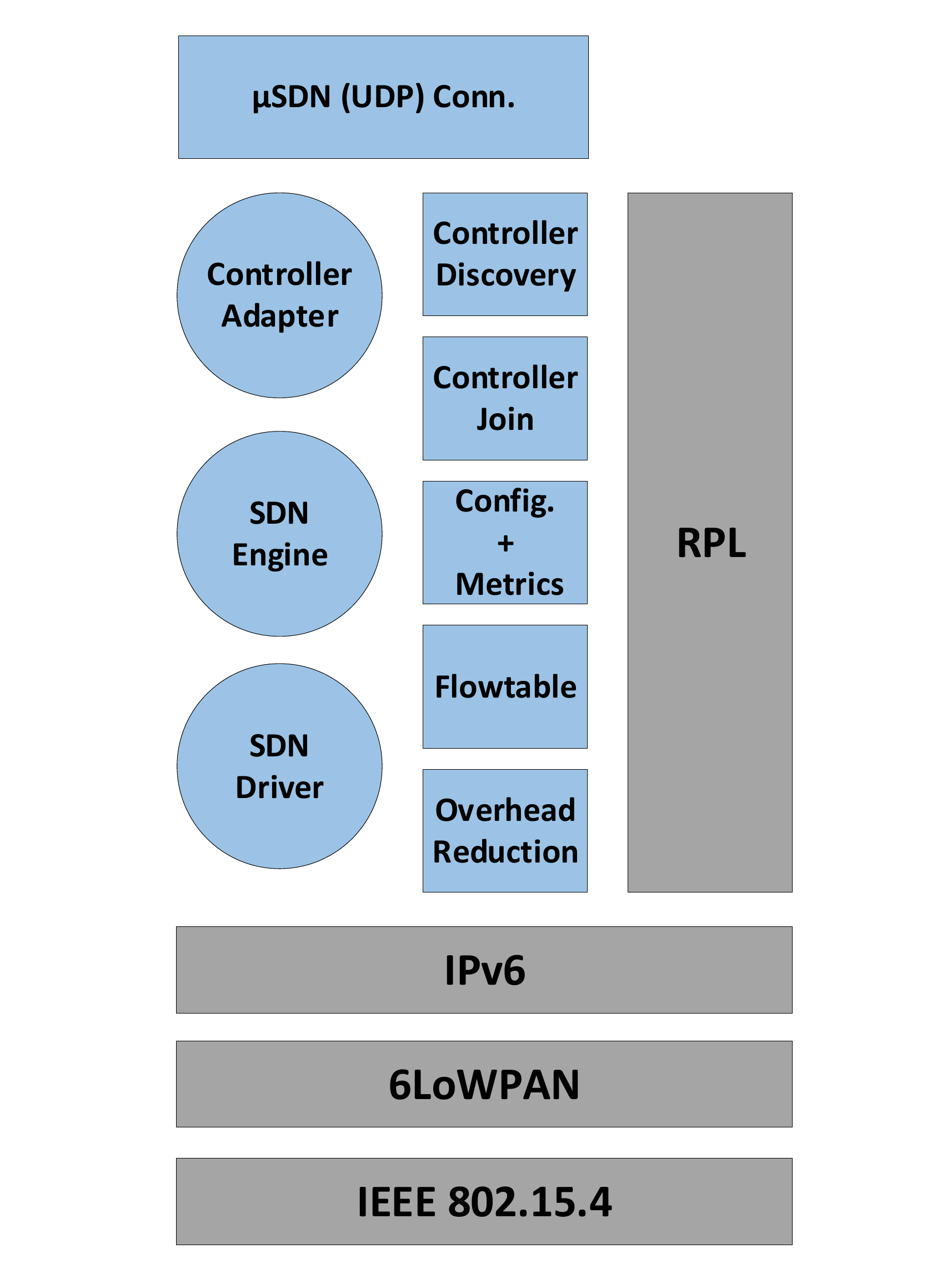}
  \caption{$\mu$SDN architecture. Blue denotes $\mu$SDN modules; whilst gray shows the core IEEE 802.15.4-2012 and 6LoWPAN layers}
\label{fig:usdn_arch}
\end{figure}

$\mu$SDN provides a modular architecture and API which allows application specific features to be separated from core SDN processes. This architecture is presented in Figure \ref{fig:usdn_arch} and is fully integrated with the IEEE 802.15.4-2012 protocol stack. It has been tested in Cooja using TI's exp5438 platform with MSP430F5438 CPU and CC2420 radio.

\subsection{Modular Stack Implementation}
The \textit{$\mu$SDN Stack} provides a layered architecture and API to separate core function handling from the specifics of the SDN implementations.

\begin{itemize}
\item \textit{$\mu$SDN Protocol}: $\mu$SDN uses its own lightweight protocol for controller communication. It's transported over UDP to allow for secure DTLS (Datagram Transport Layer Security) when communicating with controllers outside the mesh, and is highly optimized to ensure no packet fragmentation.
\item \textit{Controller Adapter:} Exposes an abstract controller interface to the SDN layer, allowing the $\mu$SDN Protocol to be switched out for any other protocol which implements that interface. 
\item \textit{SDN Engine:} Defines how the messages to and from the controller are handled. It is essentially the concrete implementation of the protocol logic, dictating how the node handles controller communication. 
\item \textit{SDN Driver:} Provides an API for the SDN Engine by defining how the flowtable is handled. It provides high-level functions for performing certain tasks through the setting of flowtable entries such as: creating firewall entries, setting routing paths through the network, or aggregating flows. It also provides handling of the flowtable actions, and determines how and when nodes should defer to the controller for instruction.
\end{itemize}

\begin{table*}[t]
	\renewcommand{\arraystretch}{1.0}
	\caption{$\mu$SDN Packet Types} 
    \label{table:usdn_packets}
	\centering
    \begin{tabular}{ |l|l|l|l| }
    \hline
      	\bfseries Packet Type & \bfseries Direction & \bfseries Behavior & \bfseries Description\\ \hline
		Node State Update (NSU) & UP & Periodic & Updates the controller with node information \\ \hline
        Flowtable Query (FTQ) & UP & Intermittent & Requests flowtable instructions from controller \\ \hline
        Flowtable Set (FTS) & DOWN & Intermittent & Sets an entry in a node's flowtable \\ \hline
        Configuration (CONF) & DOWN & Initial & Configures a node's non-flowtable settings \\
    \hline
    \end{tabular}
\end{table*}

\begin{figure*}[t]
\centering
  \begin{subfigure}{1\columnwidth}
    \centering
    \includegraphics[width=0.8\columnwidth]{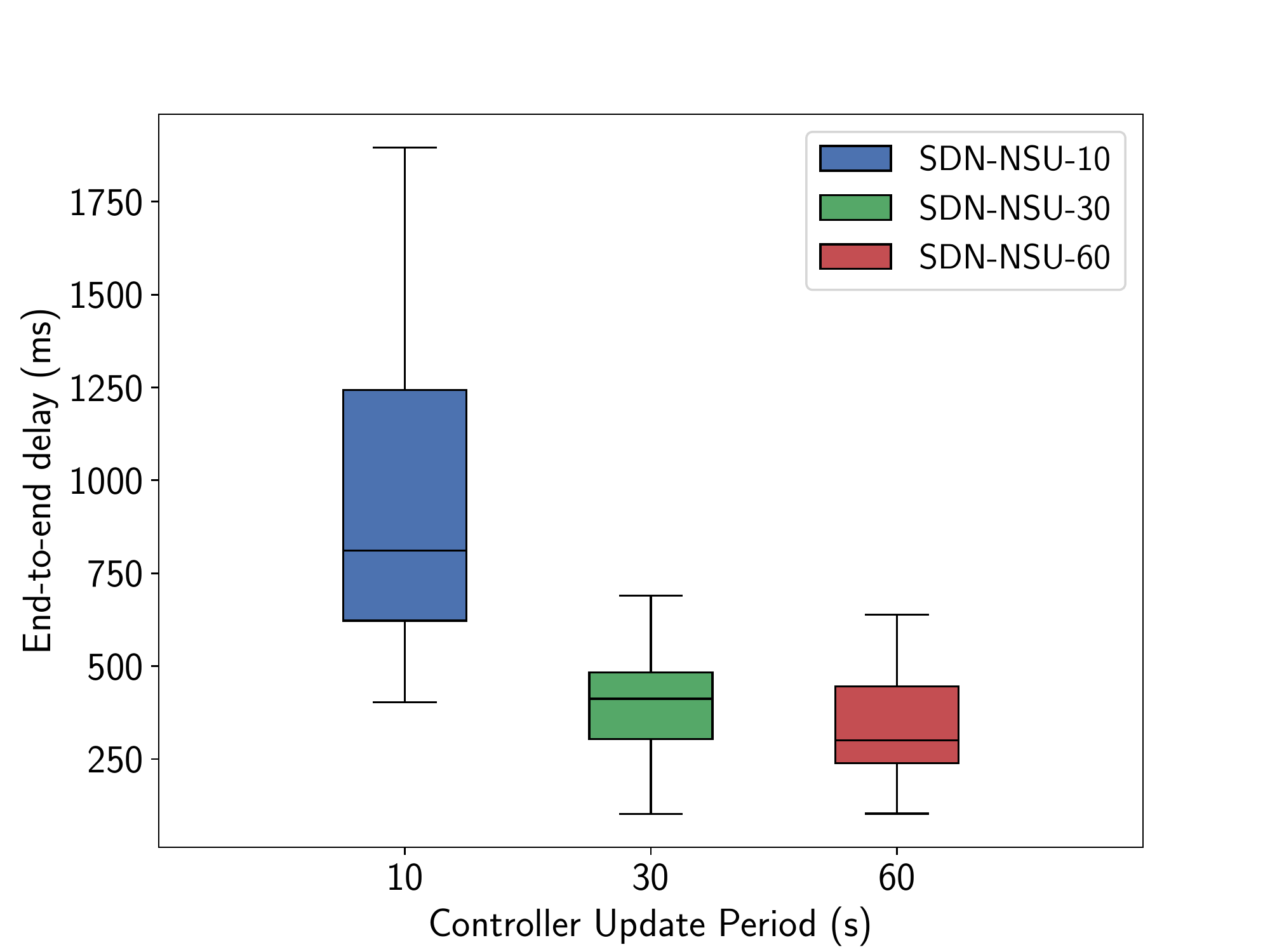}
    \caption{Effect of NSU period on average end-to-end application latency.}
    \label{fig:30_node_join}
  \end{subfigure}
  \begin{subfigure}{1\columnwidth}
    \centering
    \includegraphics[width=0.8\columnwidth]{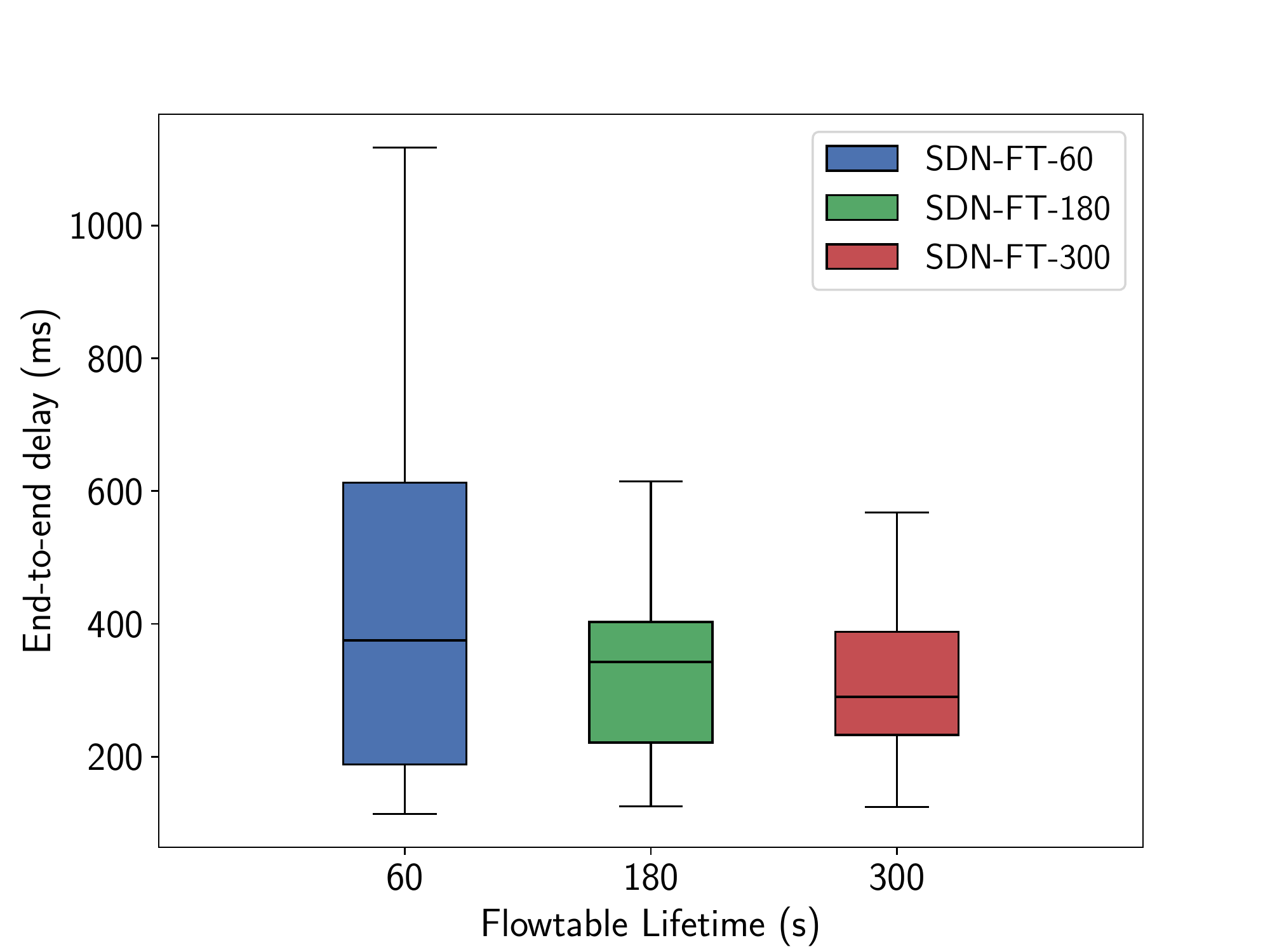}
    \caption{Effect of FT lifetime on average end-to-end application latency.}
  	\label{fig:30_node_pdr}
  \end{subfigure}
\caption{Comparison of a the effect of range of SDN controller update periods and flowtable lifetime on application traffic delay. Simulation parameters are detailed in Table \ref{table:sim_params}}
\label{fig:param_test}
\end{figure*}

\subsection{Core SDN Processes}
The \textit{$\mu$SDN Core} provides essential SDN processes, allowing protocol and framework specific implementations to be built on top.

\begin{itemize}
\item \textit{Controller Discovery:} Integrates with the network's distributed routing protocol. This gives the node \textit{fallback} or \textit{default} routing in the event that a node loses its connection to the controller. Although this is currently RPL (both Storing and Non-Storing), this could in theory be any distributed routing protocol implemented within the network. This grants controller connections within $\mu$SDN an element of robustness, and ensures nodes will always attempt to find a path to the controller. 
\item \textit{Controller Join:} This Layer-3 process employs both the underlying RPL topology, as well as the $\mu$SDN protocol provided by the SDN stack. When the controller receives a RPL DAO (destination advertisement) message, it will send a $\mu$SDN CONF message to the joining node in order to initialize the node and provide configuration information. The joining node uses this CONF message as acknowledgement that it is connected to the controller.
\item \textit{Configuration and Metrics:} Allows controllers to setup the SDN network, choose which metrics to receive from the node, and select what information to receive in controller requests.
\item \textit{Flowtable:} Optimized for memory due to node hardware constraints. Using a similar approach to Protocol Oblivious Forwarding (PoF) \cite{pof} this allows for a flowtable with a minimal memory footprint. Additionally, a Hierarchical Flowtable Structure (HFS) interface is provided to allow controllers to configure multiple flowtables with varying priority levels. This, for example, allows the controller to configure a \textit{whitelist} which is processed before the main flowtable. Packets matched in this \textit{whitelist} are then handed back to the regular Layer-3 processes.
\item \textit{Overhead Reduction:} A number of functions are implemented to mitigate SDN control overhead. CMQ \cite{cmq} is used to handle repeated flowtable misses. Partial Packet Queries (PPQ) allow flowtable requests to be sent to the controller using only partial packet information, reducing 6LoWPAN fragmentation. Source Routing Header Insertion (SRHI) allows routing headers to be inserted onto packets, and can be read by either the RPL or SDN layer. Finally, Flowtable Refresh (FR) allows controllers to instruct particularly active flowtable entries to reset their lifetimers, rather than having the entry expire.
\end{itemize}

\subsection{$\mu$SDN Protocol and Traffic Characterization}

The $\mu$SDN protocol follows the main packet types present in traditional SDN protocols such as OpenFlow: with basic flowtable request/set functionality, as well as configuration and metric update packets. All $\mu$SDN packet types are listed in Table \ref{table:usdn_packets}. As discussed in Section \ref{sec_approach}, it is essential that any SDN protocol for low-powered wireless networks is highly optimized to eliminate 6LoWPAN fragmentation, and the packets therefore have limits on the amount of information that can be sent to and from the controller. To this end, $\mu$SDN compresses information such as node addresses, as well as using node configuration tables to ensure that the controller is able to specify information sent. The traffic generated by $\mu$SDN stems from three processes: controller discovery, node updates, and requests for controller instruction.

\textbf{Controller Discovery:} $\mu$SDN employs the RPL protocol to inform the controller of nodes which have joined the DAG, and are therefore reachable. However, it generates additional traffic in the form of a Configuration (CONF) response to each node joining. This allows nodes which have joined the network to receive initialization information from the controller, such as: NSU timer settings, flowtable lifetimes, and default flowtable entries.

\textbf{Node Updates:} A Node State Update (NSU) message, from a node to the controller, carries information about that node, such as energy, node state, and buffer congestion. This includes observations about its immediate neighbors and link performance. These periodic messages are sent on a timer process within the \textit{Configuration and Metrics} module, that can be configured by the controller using a CONF message.

\textbf{Controller Instruction:}
Flowtable Query (FTQ) packets are sent from a node to the controller in response to a flowtable miss, i.e. the SDN checks the flowtable for instructions on how to handle a packet but is unable to find a matching entry. With Partial Packet Queries (PPQ), FTQ messages send a portion of the packet data up to the controller. The controller then actions that data, and transmits a response back to the sender in the form of a Flowtable Set (FTS) message. The behavior of this traffic is by nature intermittent, though it also depends on whether or not the flowtable uses source routing headers for forwarding. If source routing isn't used, then it will exhibit bursty behavior as FTQ packets are generated by each node in the path between the source and destination. 


%% file: sec_04_results.tex
\begin{figure*}[t]
\centering
  \begin{subfigure}{1\columnwidth}
    \centering
    \includegraphics[width=0.8\columnwidth]{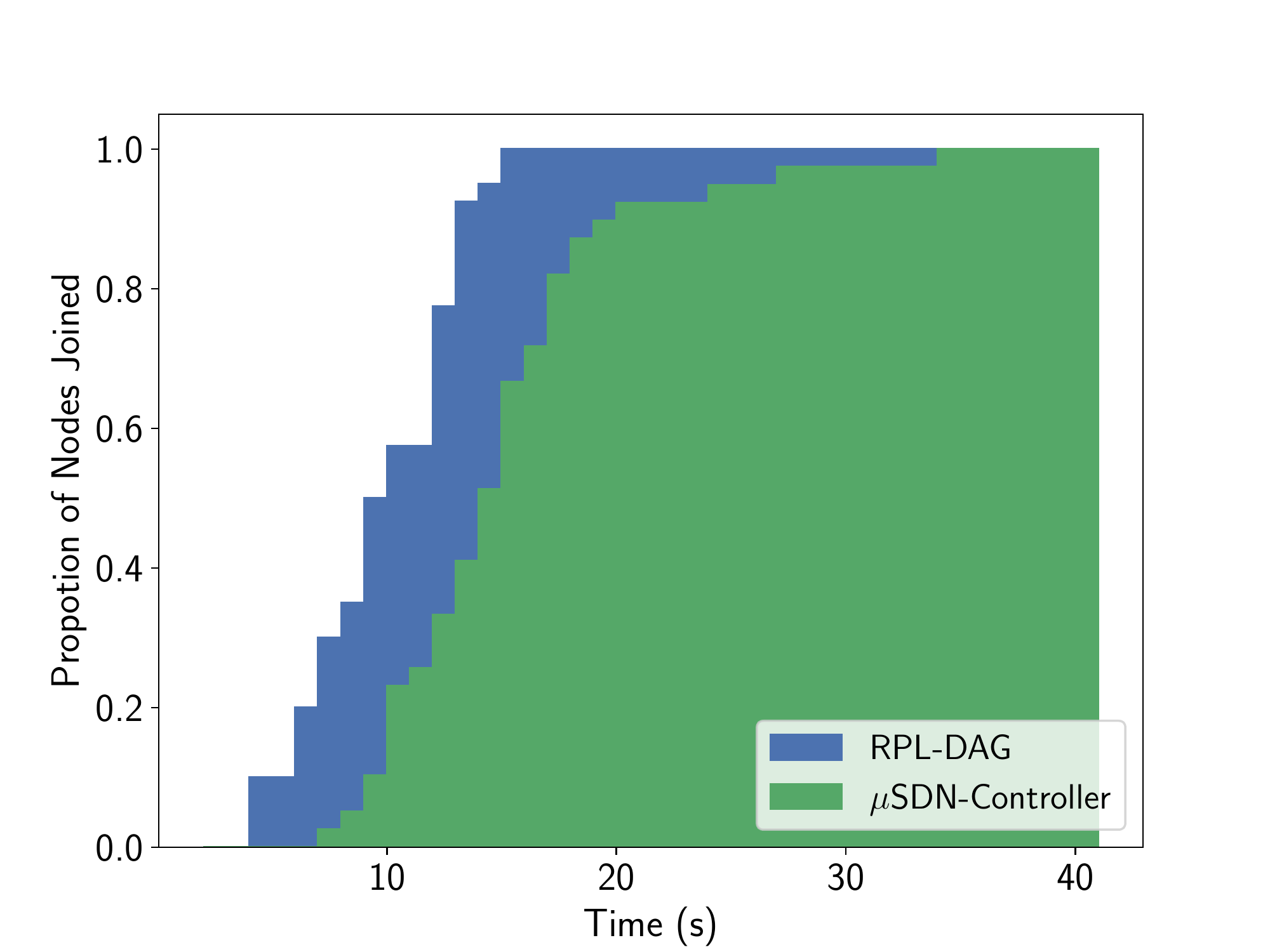}
    \caption{Node join times for both the RPL DAG and $\mu$SDN controller.}
    \label{fig:node_join}
  \end{subfigure}
  \begin{subfigure}{1\columnwidth}
    \centering
    \includegraphics[width=0.8\columnwidth]{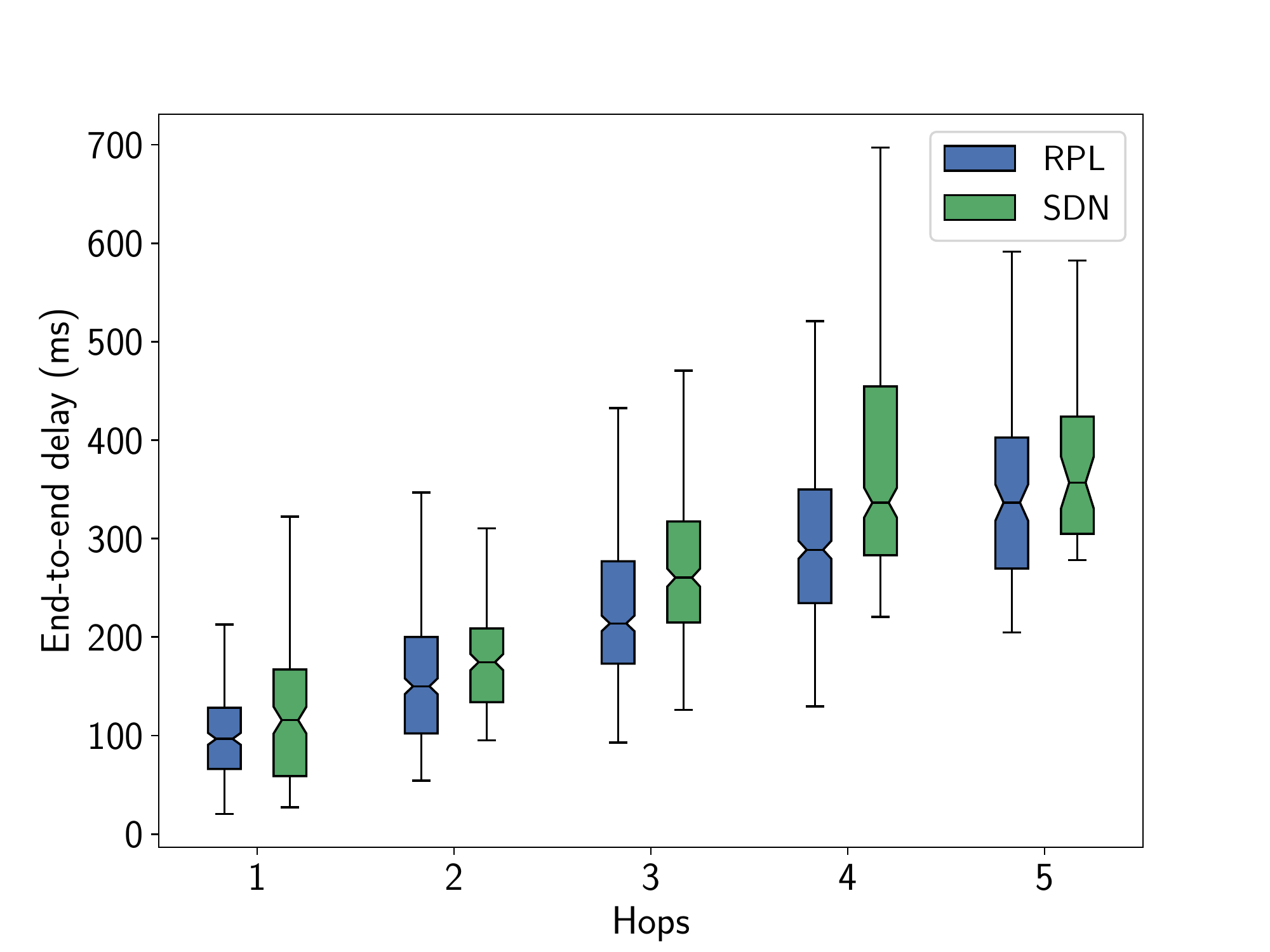}
    \caption{End-to-end application flow latency.}
    \label{fig:node_e2e}
  \end{subfigure}
  \begin{subfigure}{1\columnwidth}
    \centering
    \includegraphics[width=0.8\columnwidth]{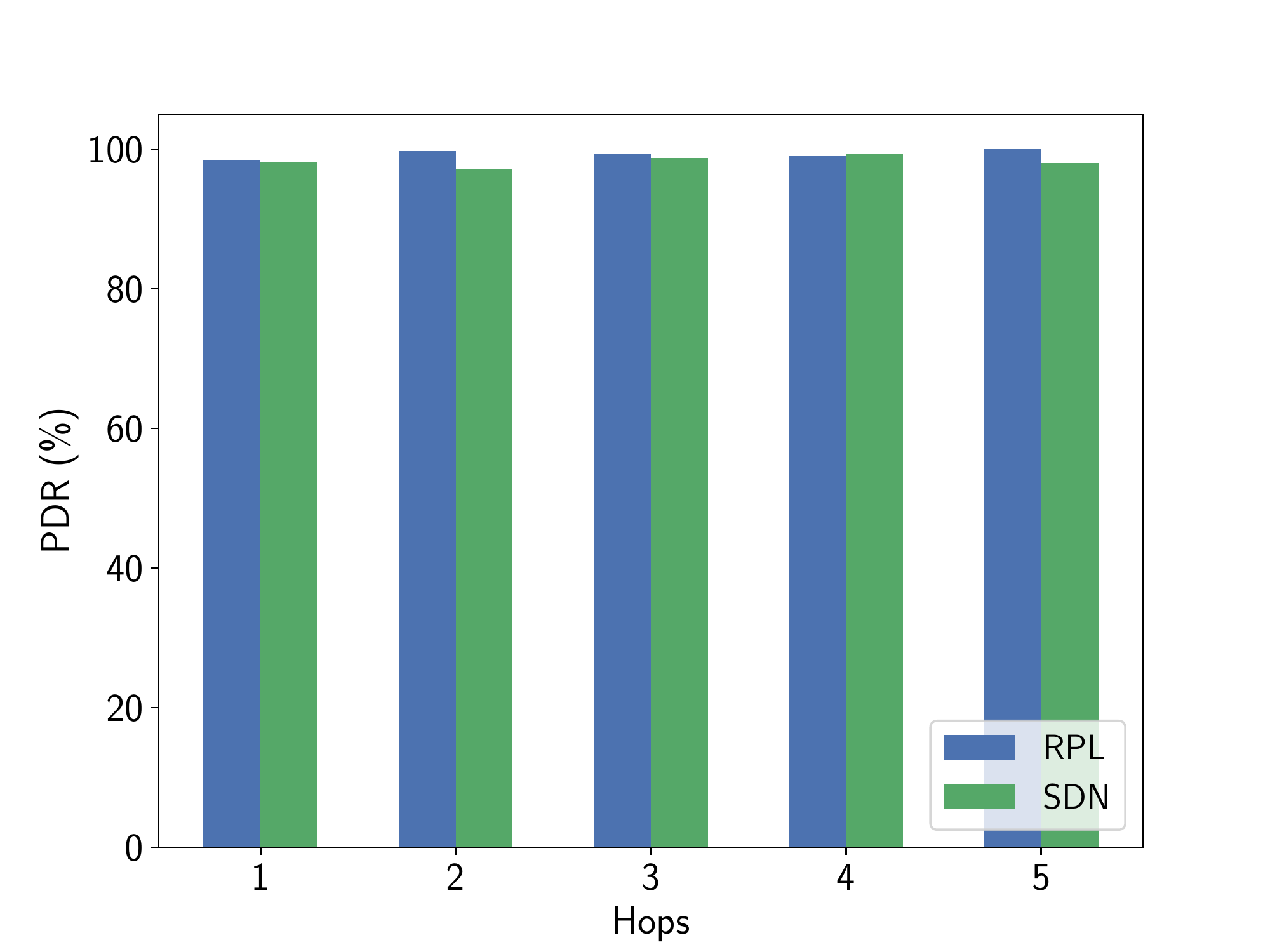}
    \caption{Packet Delivery Ratio (PDR).}
  	\label{fig:node_pdr}
  \end{subfigure}
  \begin{subfigure}{1\columnwidth}
    \centering
    \includegraphics[width=0.8\columnwidth]{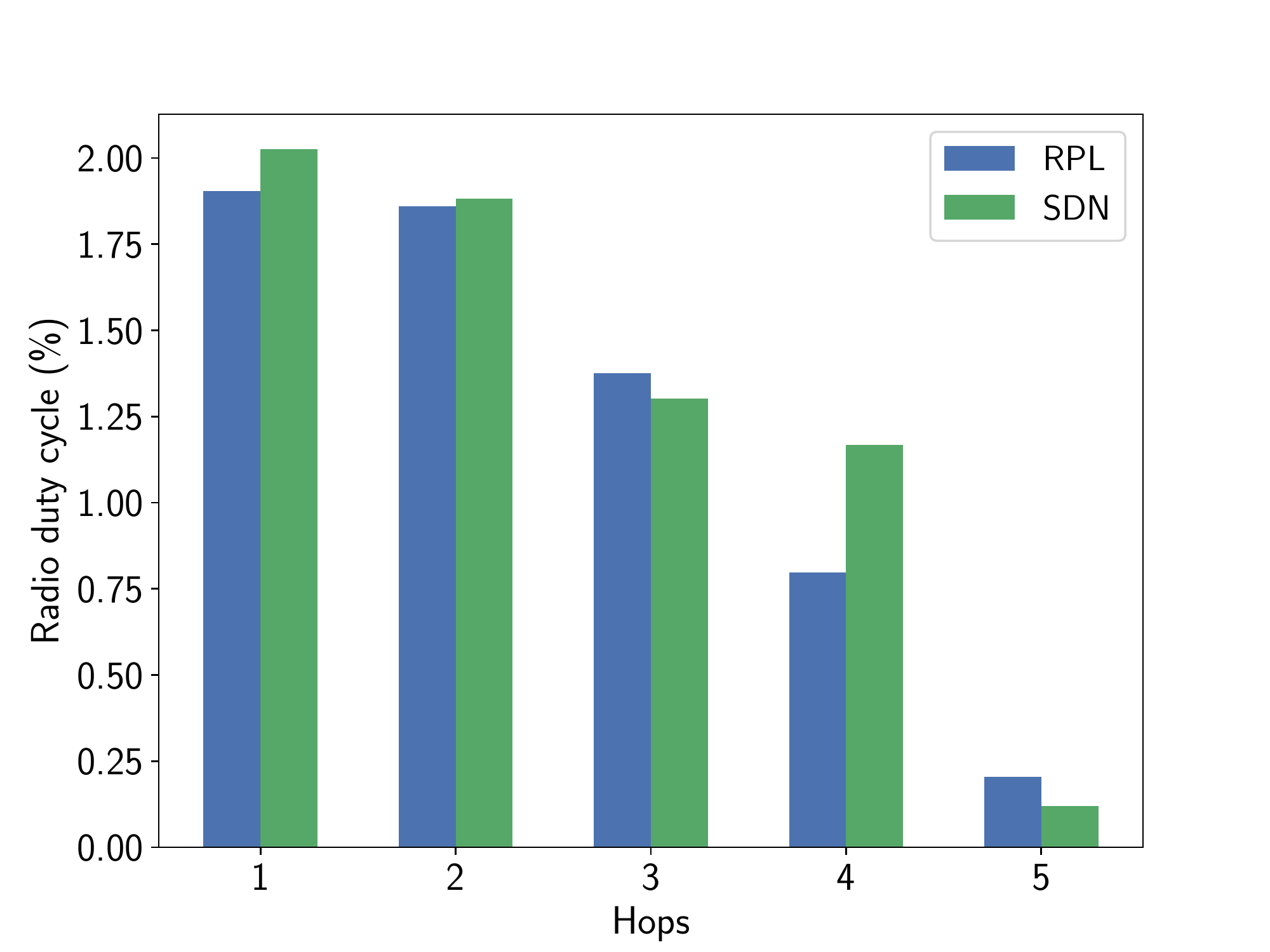}
    \caption{Radio Duty Cycling (RDC).}
 	 \label{fig:node_rdc}
  \end{subfigure}
\caption{Overall performance of $\mu$SDN in comparison to a conventional RPL network. Network consists of 30 nodes with a maximum of 5 hops to the controller or DAG root}
\label{fig:optimizd_scenario}
\end{figure*}

\section{Evaluation}
\label{sec_results}
This section evaluates $\mu$SDN against a base RPL case with no SDN implementation. Experimentation was performed on an emulated EXP5438 platform with TI's MSP430F5438 CPU and CC2420 radio, using the Cooja simulator for Contiki OS \cite{contiki}. We firstly present the scenarios, configuration, and metrics used in the evaluation. This is followed by a comparison of $\mu$SDN performance against a standard RPL network with no SDN implementation. Finally, we present a use-case scenario showing how SDN can be used within a low-power wireless network in order to programmatically manage interference and provide QoS to high-priority flows. We show that the SDN overhead can be minimized to an extent that performance is close-to, or on-par with a network that implements no SDN, and that in certain scenarios the configurability conferred by the SDN architecture can enhance network performance.

\begin{table}[ht]
	\renewcommand{\arraystretch}{1.0}
	\caption{Cooja Simulation Parameters}
    \label{table:sim_params}
	\centering
    \begin{tabular}{ |l|l| }
    \hline
      	\bfseries Parameter & \bfseries Setting \\ \hline
        Duration & 1h \\ \hline
        MAC Layer & ContikiMAC \cite{contikimac}\\ \hline
        Transmission Range & 100m \\ \hline
        Transmitting Nodes & All \\ \hline
        Receiving Node & Root/Controller \\ \hline
        Network Size & 30 Nodes \\ \hline
        Packet Send Interval & 60 - 75s \\ \hline
        Link Quality & 90\% \\ \hline
        Radio Medium & UDGM \\ \hline
        RPL Mode & Non-Storing \\ \hline
        RPL Route Lifetime & 10min \\ \hline
        RPL Default Route Lifetime & $\infty$ \\ \hline
        $\mu$SDN Update Period & 180s \\ \hline
        $\mu$SDN Flowtable Lifetime & 10min \\ 
    \hline
    \end{tabular}
\end{table}

\begin{table}[ht]
	\renewcommand{\arraystretch}{1.0}
	\caption{Interference Scenario Parameters}
    \label{table:interference_params}
	\centering
    \begin{tabular}{ |l|l| }
    \hline
      	\bfseries Parameter & \bfseries Setting \\ \hline
        Interference Period & 100ms \\ \hline
        Interference Duration & 15ms \\ \hline
        Flow $F_0$ Bit Rate & 0.25s \\ \hline
        Flow $F_1$ Bit Rate & 10s \\ 
    \hline
    \end{tabular}
\end{table}

\textbf{Configuration:}
All simulations were performed in Cooja using a Unit Disk Graph Medium (UDGM) distance loss model with the configuration specified in Table \ref{table:sim_params} (unless otherwise stated). Configuration parameters specific to the interference scenario are specified in Table \ref{table:interference_params}

\textbf{Scenarios:} 
We evaluate $\mu$SDN in the following scenarios.
\begin{itemize}
\item \textit{SDN Traffic Test:} We examine the effect of update period and flowtable lifetime settings on SDN performance.  
\item \textit{Full Overhead Reduction:} We evaluate $\mu$SDN with all overhead reduction mechanisms employed. The intent is to show a broad analysis of the effect of an optimized low-overhead SDN framework on network performance.
\item \textit{Interference Re-Route:} We demonstrate how $\mu$SDN can be used to counter interference in the network, providing reduced delay and jitter to high-priority flows.
\end{itemize}

\textbf{Metrics:} 
We discuss the following performance metrics.
\begin{itemize}
\item \textit{Node Join Times:} Time taken until all nodes have discovered both the RPL DAG and SDN controller.
\item \textit{Traffic Ratio:} Overhead incurred both by RPL and SDN, with respect to application traffic transmitted from each node in the network.
\item \textit{End-to-End Latency:} The effect of SDN overhead on application traffic delay.
\item \textit{Packet Delivery Ratio:} The effect of SDN overhead on network reliability.
\item \textit{Radio Duty Cycle (RDC):} The effect of SDN overhead on node energy. 
\end{itemize}

\subsection{Scenario: SDN Traffic Test} \label{sec_param_test}
We compare the effect of controller update periods and flowtable lifetimes on the performance of an SDN network. Figure \ref{fig:param_test} highlights how application delay is sensitive to increases in the frequency of FTQ/FTS transmissions and NSU update period.

\subsection{Scenario: Full Overhead Reduction} \label{sec_full_reduction}

We evaluate the SDN performance in Figure \ref{fig:optimizd_scenario}, where $\mu$SDN has been configured to reduce SDN overhead through SRHI, FR, CMQ, and PPQ as discussed in Section \ref{sec_design}. 

\begin{figure}[ht]
  \centering
  \includegraphics[width=0.8\columnwidth]{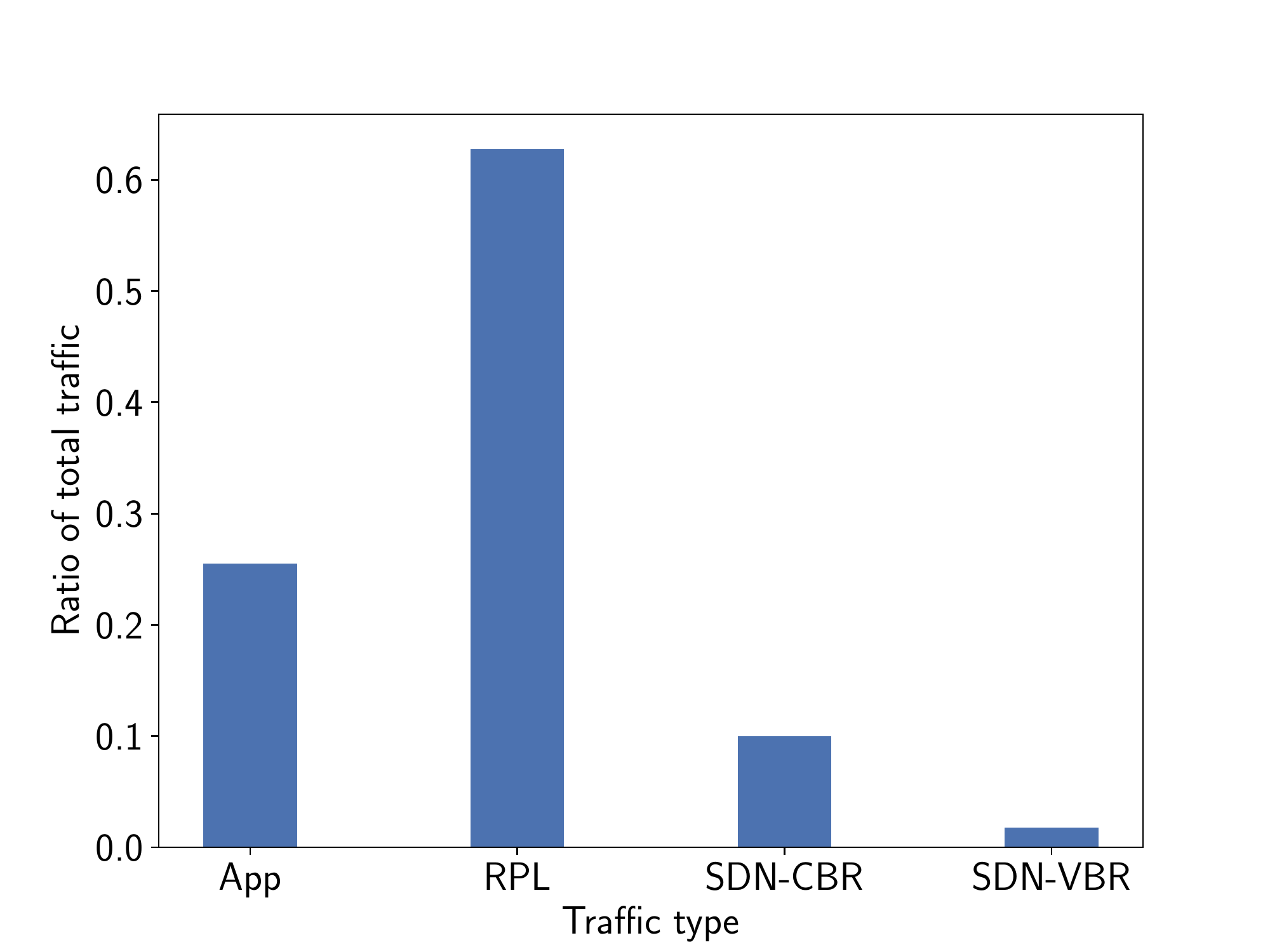}
  \caption{Ratio of application, RPL, and SDN traffic in a $\mu$SDN network.}
  \label{fig:traffic_ratio}
\end{figure}

The ratio of traffic generated in a $\mu$SDN network is examined in Figure \ref{fig:traffic_ratio}. While the RPL traffic clearly presents the highest overhead within the network, it should be noted that this is a combined figure of DIS, DIO, and DAO messages; only the latter of which is transmitted across multiple hops, while the others are exchanged between neighbors. Constant Bit Rate (periodic) and Variable Bit Rate (intermittent) SDN messages are shown separately. These are NSU and CONF/FTQ/FTS messages respectively.

In Figure \ref{fig:node_join} we present the time taken for all nodes in the network to join both the RPL DAG, and the SDN controller. In the case of the former, this is the time for the controller to learn about the routing path to that node through RPL DAO messages, which then triggers the join process.

End-to-end application latency is evaluated in Figure \ref{fig:node_e2e}. Although there is a slight increase in delay for application packets in the $\mu$SDN scenario, this is generally consistent with the slight overhead incurred by the SDN processes at each node. That is, as each node needs to perform a flowtable lookup for incoming packets, this lookup time increases the further the source node is from the destination.

Figure \ref{fig:node_pdr} shows $\mu$SDN application traffic PDR against application traffic routed through RPL. $\mu$SDN experiences a slightly lower PDR due to increased congestion and MAC-layer drops shortly after initialization. As nodes forward application packets through SRHI they need to receive this source routing header from the controller. The increased network activity means that FTQ/FTS packets are occasionally lost, and the application packet is dropped.

As $\mu$SDN operates on top of the RPL protocol there is always an associated cost, particularly when considering the energy performance of nodes. Figure \ref{fig:node_rdc} shows the average RDC of nodes in a 30 node network at 1 to 5 hops, where there is a slight increase over the RPL case.

\subsection{Scenario: Interference Re-Routing} \label{sec_interference_rr}
\begin{figure}[ht]
  \centering
  \includegraphics[width=0.7\columnwidth]{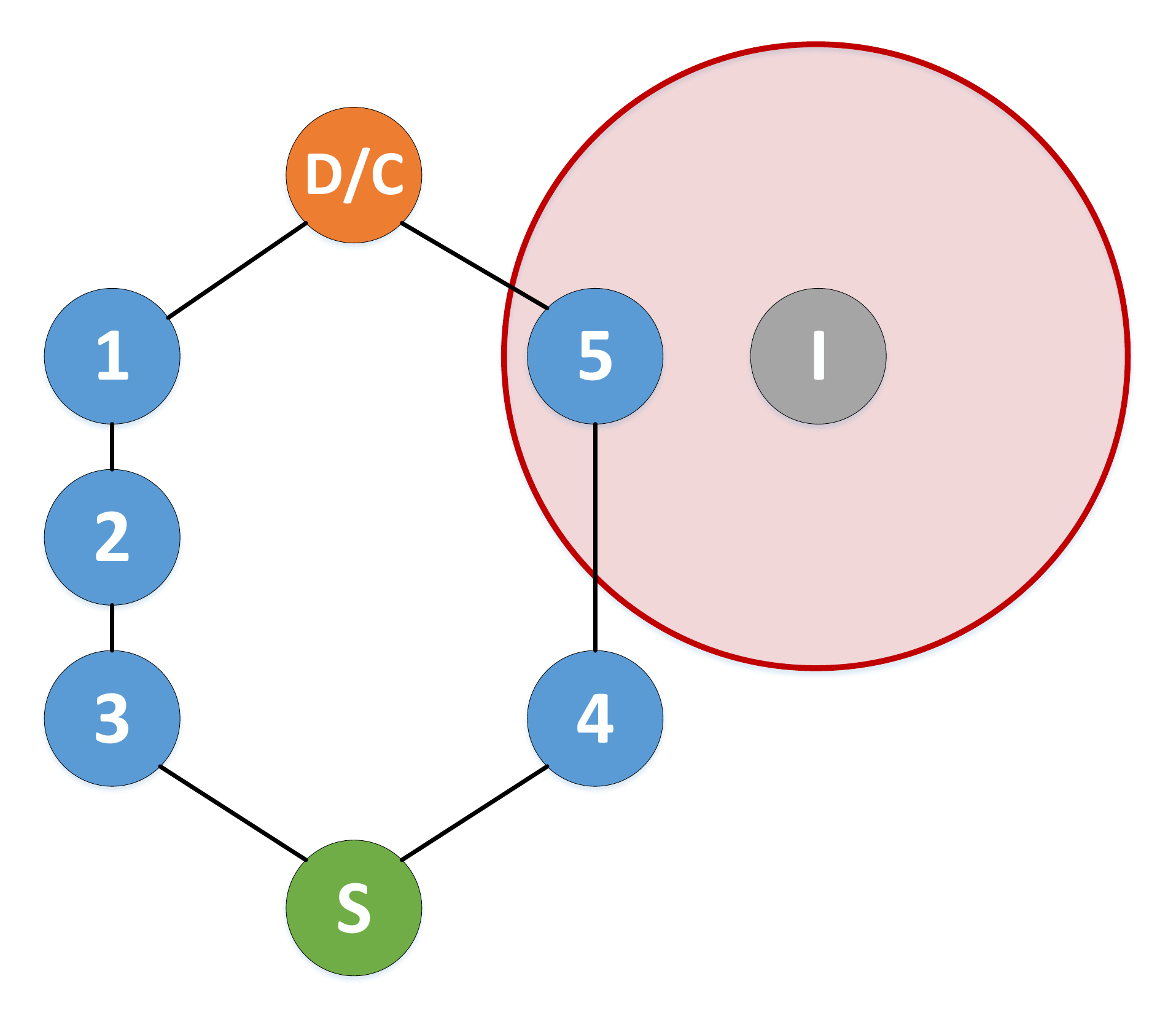}
  \caption{Topology of intermittent interference scenario. The source node (S) is in green, whilst the destination/controller node is in orange (D/C). Intermittent interference is generated at I, interfering with node 5}
\label{fig:interference-rr-topo}
\end{figure}

Though SDN inevitably adds an associated cost to general performance, the authors of this paper argue that the configurablity conferred by SDN architecture allows increased QoS in cases where a distributed protocol will fail. To the effectiveness of $\mu$SDN programmability we implemented an interference scenario as shown in Figure \ref{fig:interference-rr-topo} and Table \ref{table:interference_params}. 

\begin{figure}[ht]
  \centering
  \includegraphics[width=0.8\columnwidth]{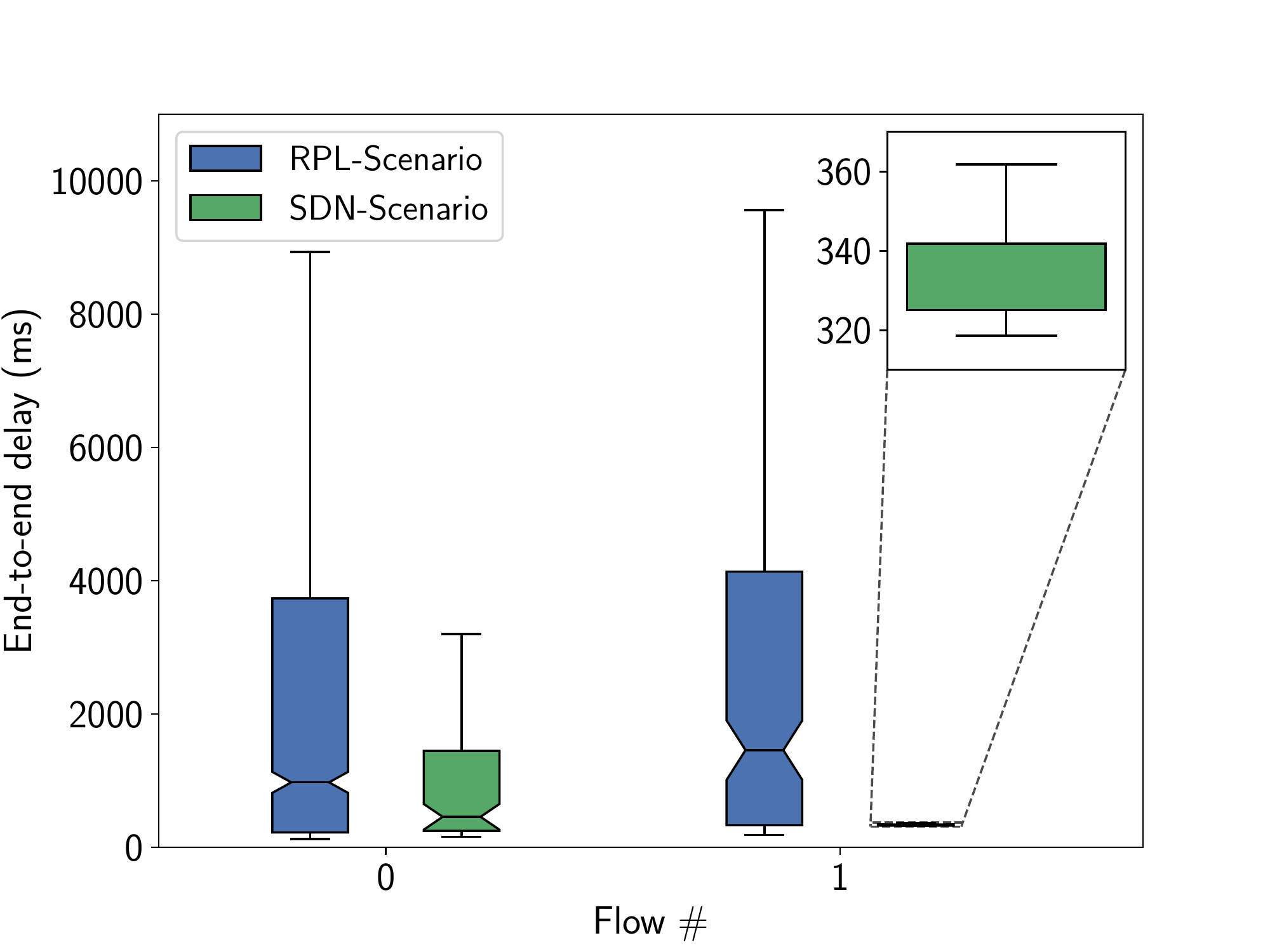}
  \caption{Delay and jitter of flows in the intermittent interference re-routing scenario (Section \ref{sec_interference_rr}). Compares a SDN scenario against a RPL scenario with no SDN. In the SDN scenario, $\mu$SDN is configured to reroute flow $F_1$ around the interference. The achieved reduction in delay and jitter can be seen in the highlighted area of the figure.}
  \label{fig:interference_contikimac}
\end{figure}

In this setup, a source node creates two flows, $F_0$ and $F_1$. $F_0$ is a low priority, but high volume flow, whereas $F_1$ is critical flow with a much lower bit rate but high priority. RPL Objective Function (OF) Zero was used, which instructs RPL nodes to choose their parents based on node rank. In this case the source node \textit{S} will receive DAG information from both node \textit{3} and node \textit{4}, however it will choose \textit{4} as it's parent as that node will have a lower rank due to its proximity to the root node \textit{D}, which in this scenario is also the destination node and the SDN controller. An interferer node was placed so that node \textit{5} would experience a short burst (15ms) of interference every 100ms, causing flows across the RPL route to experience a high degree of degradation. As the interference is not constant, the RPL DAG is unable to heal and form a new path through node \textit{3}.

The introduction of $\mu$SDN to the network allows the controller to handle flows individually, and re-route $F_1$ through \textit{3} even though it is the longer path and is not the next hop dictated by the RPL OF. Flow $F_1$ is therefore able to bypass the interference, experiencing reduced delay and jitter whilst Flow $F_0$ continues to be routed using RPL. This also has the side-effect of reducing the delay of $F_0$ as the path [S, 5, 4, D] experiences less traffic. These results are shown in Figure \ref{fig:interference_contikimac}, where $\mu$SDN exhibits dramatically reduced delay in comparison to the scenario without the benefit of SDN configurability.

%% file: sec_05_conclusion.tex
\section{Conclusions}
\label{sec_conclusion}
As low-power wireless communications move beyond simple sensor networks and towards multi-tenant and multi-application IoT scenarios, there is an increasing need for flexibility within the network. This paper has introduced $\mu$SDN, a lightweight SDN architecture which overcomes the challenges of introducing SDN in low-power wireless networks. We argue that co-existence with a distributed routing protocol is necessary to provide a framework for \textit{controller discovery }, although this means that any control traffic generated through SDN is an additional overhead. To this extent we have proposed the combination of a number of overhead reduction functions, and $\mu$SDN implements these to substantially mitigate the cost of SDN within a constrained environment. We have shown that it maintains comparable scalability with RPL-based IEEE 802.15.4-2012 networks, whilst providing the network with the opportunities inherent in SDN architecture, such as \textit{Global Knowledge}, \textit{Network (Re)Configurability}, and \textit{Virtualization}. In particular, this paper has demonstrated a scenario where $\mu$SDN is used to implement per-flow QoS handling within a simple network under intermittent interference, showing how $\mu$SDN can provide redundancy to priority flows, where we achieve considerable reduction in latency and jitter in comparison to a conventional low-power wireless network.


%% file: main.bbl
\begin{thebibliography}{10}
\providecommand{\url}[1]{#1}
\csname url@samestyle\endcsname
\providecommand{\newblock}{\relax}
\providecommand{\bibinfo}[2]{#2}
\providecommand{\BIBentrySTDinterwordspacing}{\spaceskip=0pt\relax}
\providecommand{\BIBentryALTinterwordstretchfactor}{4}
\providecommand{\BIBentryALTinterwordspacing}{\spaceskip=\fontdimen2\font plus
\BIBentryALTinterwordstretchfactor\fontdimen3\font minus
  \fontdimen4\font\relax}
\providecommand{\BIBforeignlanguage}[2]{{%
\expandafter\ifx\csname l@#1\endcsname\relax
\typeout{** WARNING: IEEEtran.bst: No hyphenation pattern has been}%
\typeout{** loaded for the language `#1'. Using the pattern for}%
\typeout{** the default language instead.}%
\else
\language=\csname l@#1\endcsname
\fi
#2}}
\providecommand{\BIBdecl}{\relax}
\BIBdecl

\bibitem{sdn_comprehensive_survey}
D.~Kreutz, F.~M.~V. Ramos, P.~E. Veríssimo, C.~E. Rothenberg, S.~Azodolmolky,
  and S.~Uhlig, ``Software-defined networking: A comprehensive survey,''
  \emph{Proceedings of the IEEE}, vol. 103, no.~1, pp. 14--76, Jan 2015.

\bibitem{6tisch_ietf_architecture}
\BIBentryALTinterwordspacing
P.~Thubert, ``{An Architecture for IPv6 over the TSCH mode of IEEE 802.15.4},''
  Internet Engineering Task Force, Internet-Draft
  draft-ietf-6tisch-architecture-11, Jan. 2017, work in Progress. [Online].
  Available:
  \url{https://datatracker.ietf.org/doc/html/draft-ietf-6tisch-architecture-11}
\BIBentrySTDinterwordspacing

\bibitem{rpl_rfc}
A.~Brandt, J.~Vasseur, J.~Hui, K.~Pister, P.~Thubert, P.~Levis, R.~Struik,
  R.~Kelsey, T.~H. Clausen, and T.~Winter, ``{RPL: IPv6 Routing Protocol for
  Low-Power and Lossy Networks},'' IETF RFC 6550, Mar. 2012.

\bibitem{contiki}
A.~Dunkels, B.~Gronvall, and T.~Voigt, ``Contiki - a lightweight and flexible
  operating system for tiny networked sensors,'' in \emph{29th Annual IEEE
  International Conference on Local Computer Networks}, Nov 2004, pp. 455--462.

\bibitem{shenker_future_past}
\BIBentryALTinterwordspacing
{Open Networking Summit}, ``The future of networking, and the past of protocols
  - scott shenker.'' [Online]. Available:
  \url{https://www.youtube.com/watch?v=YHeyuD89n1Y}
\BIBentrySTDinterwordspacing

\bibitem{openflow}
N.~McKeown, T.~Anderson, H.~Balakrishnan, G.~Parulkar, L.~Peterson, J.~Rexford,
  S.~Shenker, and J.~Turner, ``Openflow: Enabling innovation in campus
  networks,'' \emph{SIGCOMM Comput. Commun. Rev.}, vol.~38, no.~2, pp. 69--74,
  Mar. 2008.

\bibitem{6lowpan_rfc}
P.~Thubert and J.~Hui, ``{Compression Format for IPv6 Datagrams over IEEE
  802.15.4-Based Networks},'' IETF RFC 6282, Sep. 2011.

\bibitem{wsdn_survey_taxonomy}
I.~T. Haque and N.~Abu-Ghazaleh, ``Wireless software defined networking: A
  survey and taxonomy,'' \emph{IEEE Communications Surveys Tutorials}, vol.~18,
  no.~4, pp. 2713--2737, Fourthquarter 2016.

\bibitem{sdwn_opportunities_challenges}
K.~Sood, S.~Yu, and Y.~Xiang, ``Software-defined wireless networking
  opportunities and challenges for internet-of-things: A review,'' \emph{IEEE
  Internet of Things Journal}, vol.~3, no.~4, pp. 453--463, Aug 2016.

\bibitem{sdn_for_iot_survey}
S.~Bera, S.~Misra, and A.~V. Vasilakos, ``Software-defined networking for
  internet of things: A survey,'' \emph{IEEE Internet of Things Journal},
  vol.~PP, no.~99, pp. 1--1, 2017.

\bibitem{sensor_openflow}
T.~Luo, H.~P. Tan, and T.~Q.~S. Quek, ``Sensor openflow: Enabling
  software-defined wireless sensor networks,'' \emph{IEEE Communications
  Letters}, vol.~16, no.~11, pp. 1896--1899, November 2012.

\bibitem{cmq}
T.~Luo, H.~P. Tan, P.~C. Quan, Y.~W. Law, and J.~Jin, ``Enhancing
  responsiveness and scalability for openflow networks via control-message
  quenching,'' in \emph{2012 International Conference on ICT Convergence
  (ICTC)}, Oct 2012, pp. 348--353.

\bibitem{sdwn}
S.~Costanzo, L.~Galluccio, G.~Morabito, and S.~Palazzo, ``Software defined
  wireless networks: Unbridling sdns,'' in \emph{2012 European Workshop on
  Software Defined Networking}, Oct 2012, pp. 1--6.

\bibitem{pof}
H.~Song, ``Protocol-oblivious forwarding: Unleash the power of sdn through a
  future-proof forwarding plane,'' in \emph{Proceedings of the Second ACM
  SIGCOMM Workshop on Hot Topics in Software Defined Networking}, ser. HotSDN
  '13.\hskip 1em plus 0.5em minus 0.4em\relax New York, NY, USA: ACM, 2013, pp.
  127--132.

\bibitem{sdn_wise}
L.~Galluccio, S.~Milardo, G.~Morabito, and S.~Palazzo, ``Sdn-wise: Design,
  prototyping and experimentation of a stateful sdn solution for wireless
  sensor networks,'' in \emph{2015 IEEE Conference on Computer Communications
  (INFOCOM)}, April 2015, pp. 513--521.

\bibitem{coral-demo}
G.~Violettas, T.~Theodorou, S.~Petridou, A.~Tsioukas, and L.~Mamatas, ``Demo
  abstract: An experimentation facility enabling flexible network control for
  the internet of things,'' in \emph{2017 IEEE Conference on Computer
  Communications Workshops (INFOCOM WKSHPS)}, May 2017, pp. 992--993.

\bibitem{contikimac}
A.~Dunkels, ``The contikimac radio duty cycling protocol,'' 2011.

\end{thebibliography}
